\documentclass[conference]{IEEEtran}
\usepackage{cite}
\usepackage{amsmath,amssymb,amsfonts}
\usepackage{algorithmic}
\usepackage{graphicx}
\usepackage{tabularx, booktabs}
\usepackage[table]{xcolor}
\usepackage{float}
\usepackage{tikz}
\usetikzlibrary{shapes, arrows.meta, positioning, calc}
\usepackage{caption}
\begin{document}

\title{Jamming Attacks on the Random Access Channel in 5G and B5G Networks}
\author{\IEEEauthorblockN{
        Wilfrid Azariah\IEEEauthorrefmark{1},
        Yi-Quan Chen\IEEEauthorrefmark{1},
        Zhong-Xin You\IEEEauthorrefmark{1},
        Ray-Guang Cheng\IEEEauthorrefmark{1}
        Shiann-Tsong Sheu\IEEEauthorrefmark{2},
        Binbin Chen \IEEEauthorrefmark{3}
       \\
  }
 \IEEEauthorblockA{\IEEEauthorrefmark{1}
National Taiwan University of Science and Technology, Taiwan\\
  }
 \IEEEauthorblockA{\IEEEauthorrefmark{2}
National Central University, Taiwan  \\
  }
 \IEEEauthorblockA{\IEEEauthorrefmark{3}
Singapore University of Technology and Design, Singapore \\
  }
}

\maketitle

\begin{abstract}
Random Access Channel (RACH) jamming poses a critical security threat to 5G and beyond (B5G) networks. This paper presents an analytical model for predicting the impact of Msg1 jamming attacks on RACH performance. We use the OpenAirInterface (OAI) open-source user equipment (UE) to implement a Msg1 jamming attacker. Over-the-air experiments validate the accuracy of the proposed analytical model. The results show that low-power and stealthy Msg1 jamming can effectively block legitimate UE access in 5G/B5G systems.

\end{abstract}

\begin{IEEEkeywords}
jamming attack, random access channel (RACH), OpenAirInterface (OAI)
\end{IEEEkeywords}

\section{Introduction}

\begin{figure}
\centering
\begin{tikzpicture}[auto, thick, >=Latex, scale=0.70, transform shape]

\tikzstyle{node_style} = [draw=black, fill=gray!10, rounded corners,
                          minimum width=1.cm, minimum height=0.7cm, align=center]
\tikzstyle{msg_arrow} = [->, thick]

\begin{scope}[xshift=0cm]

\node [node_style] (ue1) at (0,0) {UE};
\node [node_style] (gnb1) at (4,0) {gNB};

\draw [thick, gray!50] (ue1) -- ++(0,-5);
\draw [thick, gray!50] (gnb1) -- ++(0,-5);

\draw [msg_arrow, blue!80!black] (0,-0.8) -- node[midway, above, sloped, font=\scriptsize]{Preamble (Msg1)} (4,-1.5);
\draw [msg_arrow, red!80!black] (4,-2.0) -- node[midway, above, sloped, font=\scriptsize]{RAR (Msg2)} (0,-2.7);
\draw [msg_arrow, blue!80!black] (0,-3.2) -- node[midway, above, sloped, font=\scriptsize]{Data (Msg3)} (4,-3.9);
\draw [msg_arrow, red!80!black] (4,-4.4) -- node[midway, above, sloped, font=\scriptsize]{ACK (Msg4)} (0,-5.1);

\node at (2,-5.8) {\textbf{(a) four-step RA scheme}};

\end{scope}

\begin{scope}[xshift=6.5cm]

\node [node_style] (ue2) at (0,0) {UE};
\node [node_style] (gnb2) at (4,0) {gNB};

\draw [thick, gray!50] (ue2) -- ++(0,-5);
\draw [thick, gray!50] (gnb2) -- ++(0,-5);

\draw [msg_arrow, blue!80!black] (0,-1.5) -- node[midway, above, sloped, font=\scriptsize]{MsgA (Preamble+Data)} (4,-2.2);
\draw [msg_arrow, red!80!black] (4,-3.5) -- node[midway, above, sloped, font=\scriptsize]{MsgB (ACK)} (0,-4.2);

\node at (2,-5.8) {\textbf{(b) two-step RA scheme}};

\end{scope}

\end{tikzpicture}
\caption{Four-step and two-step RA schemes}
\label{fig:ra_procedures}
\end{figure}
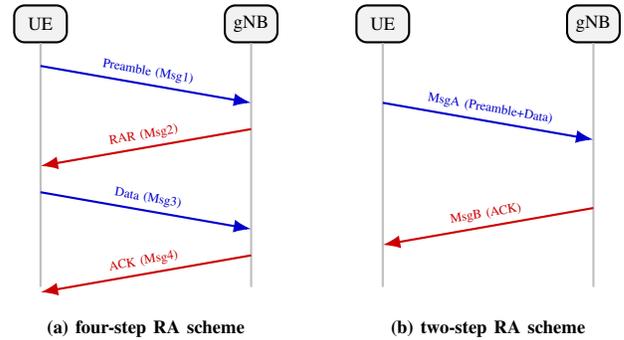

The Random Access Channel (RACH) is a physical uplink channel used in the random access (RA) procedure to enable user equipment (UE) to initiate network access and synchronization in 5G/B5G systems~\cite{TS138300,TR138864}.
In 3GPP systems, the RA procedure can be implemented using either a four-step or a two-step scheme, as shown in Fig.~\ref{fig:ra_procedures}. In the four-step RA scheme, the UE transmits a randomly selected preamble (Msg1) during a random access opportunity (RO) on the RACH. Upon successful detection of Msg1, the gNB responds with a random access response (RAR), also referred to as Msg2, carrying a timing advance, an uplink grant, and a temporary UE identification. The UE then transmits a scheduled uplink message (Msg3) using the granted resources. Finally, the gNB completes the RA procedure by transmitting a contention resolution message (Msg4). In the two-step RA scheme, the UE transmits the preamble together with the uplink message (MsgA) on the RACH. Upon successful reception of MsgA, the gNB responds with MsgB. In both cases, successful preamble detection is a critical step in completing the random access procedure. The contention-based nature of the RACH makes it vulnerable to jamming attacks~\cite{oranWG11}. In particular, jamming that targets preamble reception can directly degrade Msg1 detection. The detection of Msg1 relies on correlation-based peak detection with thresholding~\cite{sesia2011lte,figueiredo2013cacfar}. An increased threshold may prevent the gNB from detecting legitimate UE preambles, thereby reducing the RACH access success probability. In contrast, lowering the threshold increases the false alarm probability and may lead to false preamble detections and unnecessary resource allocation.

RACH jamming has been recognized as a denial-of-service (DoS) vector since the Global System for Mobile Communications (GSM) and Long Term Evolution (LTE) era. 
Early studies demonstrated that repeated RACH bursts could effectively block the channel access for legitimate UEs \cite{smartPrachJamReview}. With the evolution to 5G New Radio (NR), this vulnerability persists, as the contention-based design of the conventional 4-step RA procedure still relies on unprotected Msg1 preambles \cite{smartPrachJam}.
Although the 2-step RA procedure was introduced to reduce access latency \cite{2stepRA}, recent work shows that it may increase collision probability, particularly in dense deployment scenarios \cite{onContentionBased}. MsgA in the 2-step RA procedure still relies on a preamble-based access mechanism; hence, it remains vulnerable to jamming attacks.

Existing studies on RACH jamming focused on the feasibility of implementation or evaluation based on simulation ~\cite{smartPrachJam,implementJamAttack}. The quantitative relationship between attacker transmissions, noise threshold evolution, and access success probability remains underexplored.  This paper addresses this gap with the following contributions.
\begin{enumerate}
    \item We develop an analytical model of the evolution of the Msg1 noise threshold in the gNB, linking the attacker parameters to the access success probability of UE.
    \item We implement a protocol-aware attacker on an OAI-based testbed using Universal Software Radio Peripheral (USRP) hardware, alongside a commercial UE and gNB.
    \item We validate the analytical model experimentally, showing how the periodicity of the transmission of the attacker directly influences the access success probability of UE.
\end{enumerate}
In this paper, we present an analytical model to investigate the Msg1 threshold dynamics and validate the proposed model on an OAI-based 5G testbed. We use the model to evaluate both persistent and periodic jamming behaviors without assuming specific attacker deployment densities. The rest of the paper is organized as follows. The system model is defined in Sec. II. The analytical model is given in Sec. III.
Sec. IV shows the experimental results. Conclusion and future works are summarized in Sec. V.

\section{System Model}
\begin{figure}
\centering
\includegraphics[width=0.45\textwidth]{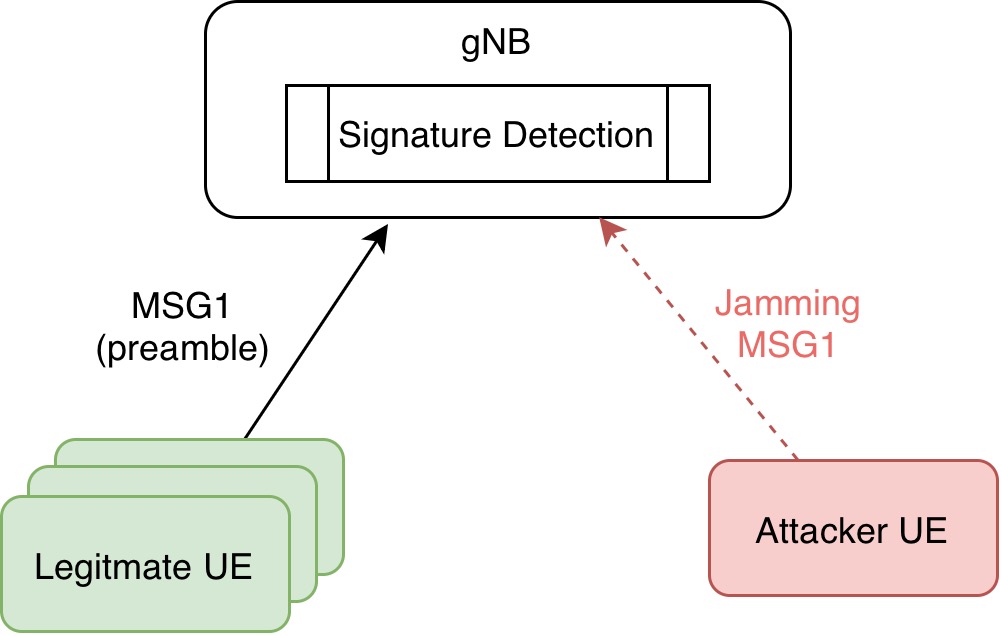}
\caption{System architecture considered in this paper}
\label{fig:system_architecture}
\end{figure}

Figure~\ref{fig:system_architecture} illustrates the system architecture considered in this paper. The system consists of three main entities: the gNB, one or more legitimate UEs, and an attacker UE. The legitimate UEs initiate the RA procedure by transmitting Msg1 preambles to the gNB. The gNB manages the RACH process at the physical layer and performs signature detection. Meanwhile, the attacker UE attempts to disrupt this process by injecting its own Msg1 signals. In practice, the gNB may employ either a fixed or adaptive noise threshold for Msg1 detection. The threshold is recursively updated using a weighted combination of current and past received power measurements together with the historical threshold value. 
The access success probability of a legitimate UE is defined as the probability that its Msg1 exceeds the gNB noise threshold and is successfully detected by the gNB. The legitimate UE initiates network access by transmitting a Msg1 preamble. An attacker may periodically transmit Msg1 preambles at a constant (maximum) power to influence the gNB’s detection threshold. By periodically injecting Msg1 transmissions, the attacker may gradually raise the threshold over time and thus, reduces the detection probability of legitimate UE.

It is assumed that the attacker can receive the Master Information Block (MIB) and System Information Block Type 1 (SIB1) from the gNB and synchronize with the cell. From SIB1, the attacker can obtain the PRACH configuration and the allocations of the ROs.
We assume that the attacker’s transmit power is significantly higher than the noise level to represent an effective jamming signal.
The initial transmit power of the legitimate UE, however, is determined through open-loop power control. At the gNB, the initial detection threshold is set to a baseline value derived from the background noise and a detection margin. The detection margin is selected based on the gNB's sensitivity requirements to balance the trade-off between miss detection and false alarm rates. Hence, an Msg1 from the legitimate UE will be processed if its received power exceeds the threshold plus a margin.

The main objective of this paper is to characterize how the gNB’s noise threshold evolves under attack and how this evolution impacts the Msg1 access probability. Specifically, we study the recursive update of the threshold based on the weighting factors and the activity of the attacker. We then investigate the effect of attacker parameters, including transmit power and periodicity, on the probability of successful Msg1 detection.

\section{Analytical Model}
\label{sec:msg1Model}

In this section, we present an analytical model to predict the impact of Msg1 jamming attacks on RACH performance. For a successful Msg1 detection, the received Msg1 power of a legitimate UE must exceed the noise threshold of the gNB calculated plus a predefined detection margin set by the gNB. Hence, the probability of a successful Msg1 detection at the $i$-th random access occasion (RO) is given by
\begin{equation}
P_{S,i} =
\begin{cases} 
1, & \text{if } p_{UE} > (p_{th,i} + \delta) \\ 
0, & \text{otherwise,} 
\end{cases}
\label{eq:success}
\end{equation}
where $p_{UE}$ is the received power metric at the gNB; $p_{th,i}$ is gNB noise threshold measured at the $i$-th RO; and $\delta$ is a predefined detection margin set by the gNB.

In practical implementations, the gNB may dynamically update the noise threshold based on current and past measurement results. In general, the detection threshold update rule is given by
\begin{equation}
p_{th,i} = \alpha \cdot p_{measured,i} + \beta \cdot p_{measured,i-1} + \gamma \cdot p_{th,i-1},
\label{eq:pth}
\end{equation}
where $p_{measured,i}$ denotes the received signal power measured at the $i$-th RO; $\alpha$ and $\beta$ are weighting factors for the current and previous measurement results, respectively; and $\gamma$ is the forgetting factor applied to the previous threshold ($\alpha$, $\beta$, $\gamma > 0$). This recursion reflects different implementations. For example, srsRAN adopts $(\alpha = 1, \beta = 0, \gamma = 0)$ (instantaneous update), while OAI uses $(\alpha = 0, \gamma = 1-\beta)$ (recursive averaging)~\cite{srsranPaper,oaiPaper}.

Let $T_a$ be the Msg1 transmission period of the attacker (i.e., the attacker transmits a Msg1 at every $T_a$-th RO). From Eq.~(\ref{eq:pth}), we can have a closed-form expression to predict the long-term behavior of the noise threshold. For a continuous attacker ($T_a = 1$) which transmits Msg1 in a constant power, we can have 
\begin{equation}
p_{measured,i} = p_{attacker}, 
\end{equation}
where $p_{attacker}$ is the received Msg1 power metric from the attacker at the gNB.
Hence, the evolution of the noise threshold follows a geometric series,
\begin{equation}
   p_{th,i} = \gamma^i p_{th,0} + (\alpha + \beta) p_{attacker} \frac{1 - \gamma^i}{1 - \gamma}.
    \label{eq:closed_form}
\end{equation}
Eq.~\ref{eq:closed_form} explicitly shows how the forgetting factor $\gamma$ determines the convergence speed of the jamming impact. As $i \to \infty$, the threshold converges to a steady-state value
\begin{equation}
p_{th,\infty} = \frac{(\alpha + \beta) p_{attacker}}{1 - \gamma}, \mbox{if \ } 0 < \gamma < 1.
\end{equation}

\begin{figure*}
\centering
\includegraphics[width=0.8\textwidth]{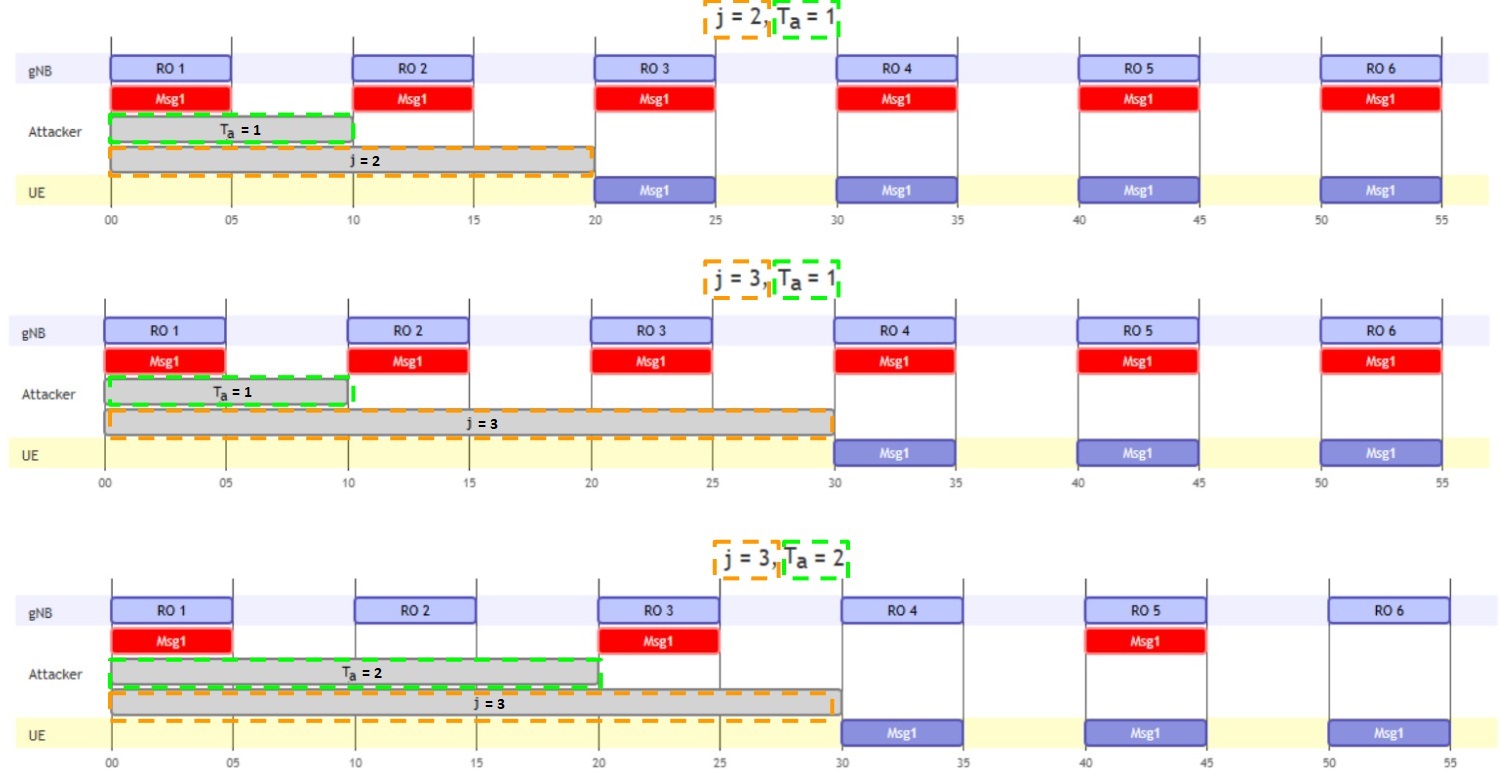}
\caption{Different attacker early start and period parameter values}
\label{fig:msg1AttackParam}
\end{figure*}

Let $p_{noise}$ is background noise level. We assumed that the received Msg1 power metric from the attacker at the gNB is higher than that of the background noise level ($p_{attacker}>p_{noise}$). Before the legitimate UE starts the RA procedure, the noise threshold is updated by
\begin{equation}
p_{measured,i} =
\begin{cases} 
p_{attacker}, & \text{if } i > 0 \text{ and } (i-1) \bmod T_a = 0 \\ 
p_{noise}, & \text{otherwise.}
\end{cases}
\label{eq:pmeasured}
\end{equation}
In this model, we focus only on the first Msg1 attempt and do not consider the power ramping or backoff of the legitimate UE. However, the same formulation can be extended to retransmissions: a second (or later) Msg1 attempt can be modeled by updating $p_{UE}$ with the ramped transmit power, and the overall access success probability can be expressed as the product of the individual success probabilities in consecutive attempts. Fig.~\ref{fig:msg1AttackParam} illustrates the different attacker early start and period parameter values. These parameters collectively define the severity of the attack and can be tuned to evaluate different jamming conditions.

From Eqs.~(\ref{eq:pth}) and (\ref{eq:pmeasured}), the attacker periodically injects power into the recursion, raising $p_{th,i}$ above natural noise levels. The rate of this increase depends on $p_{attacker}$, the update factors $(\alpha, \beta, \gamma)$, and $T_a$. A smaller $T_a$ (more frequent transmissions) accelerates threshold buildup, while a larger $T_a$ allows $p_{th,i}$ to decay toward background noise.

From Eqs.~(\ref{eq:success})–(\ref{eq:pmeasured}), the access success probability depends on the the received Msg1 power metric from the legitimate UE, the gNB noise threshold dynamics, and the attacker’s transmission periodicity. In particular:
\begin{itemize}
    \item A higher $p_{attacker}$ or a smaller $T_a$ increases $p_{th,i}$, thereby reducing $P_{S,i}$.
    \item A lower $\alpha$ or a higher $\gamma$ slows down the threshold adaptation, temporarily allowing legitimate UEs to succeed before the attack becomes dominant.
    \item A larger $p_{UE}$ relative to ($p_{th,i}+\delta$) improves the probability of successful Msg1 detection.
\end{itemize}

This analytical formulation provides the basis for predicting the degradation of network access under Msg1 jamming.

\section{Experimental Results}

\subsection{5G Testbed Setup}
\begin{figure}[h]
\centering
\includegraphics[width=0.49\textwidth]{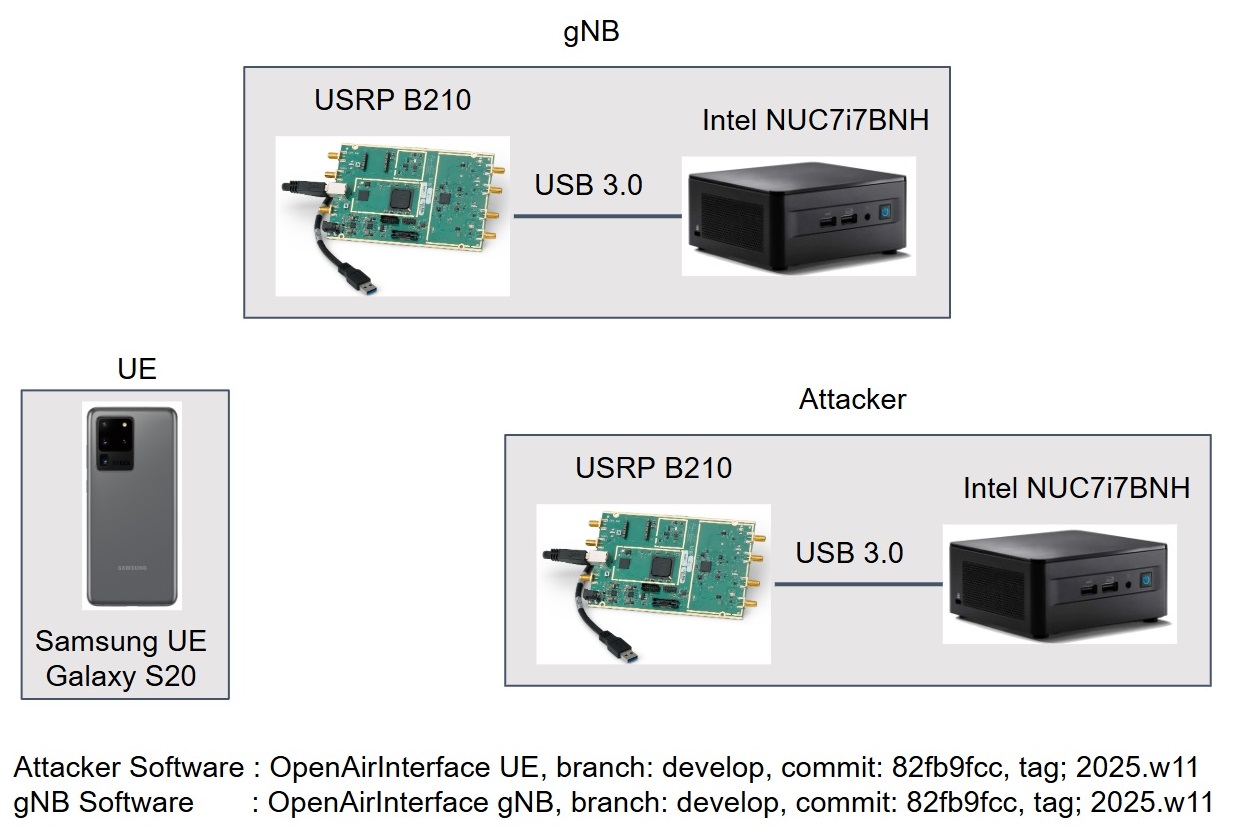}
    \caption{Testbed used in this thesis for experiment}
    \label{fig:testbed}
\end{figure}

Figure~\ref{fig:testbed} illustrates the 5G testbed to evaluate the impact of Msg1-based jamming using OAI and USRP B210 devices. Each node is deployed on an Intel NUC mini-PC (Intel Core i7-7567U, 8GB RAM, Ubuntu 22.04 LTS). The testbed consists of three components:
\begin{itemize}
\item \textbf{gNB:} Runs OAI in FR1, manages UE access, and updates the RACH noise threshold according to Msg1 detections.
\item \textbf{UE:} A commercial Samsung 5G smartphone with programmable SIM, performing real random access attempts under jamming.
\item \textbf{Attacker:} An OAI-based nrUE modified to periodically transmit Msg1 preambles, raising the gNB’s noise threshold and blocking legitimate UE access.
\end{itemize}

The experimental KPIs are collected directly from gNB console logs, which report Msg1 detections, Msg2 responses, and random access success/failure. It allows a direct comparison with the analytical model.  

\subsection{Experimental Parameters}
We follow the 3GPP specifications \cite{TS138211} to set the RACH configuration for the experiments. The selected RACH configuration parameters are summarized in Table~\ref{tab:RACHParameter}. We used hardcoded $prach-configurationIndex$ to force the attacker to align Msg1 transmissions with specific ROs. Additional parameters used in the experiments are summarized in Table~\ref{tab:experimentParameter}. Note that the power used in Table~\ref{tab:experimentParameter} refers to the power measured by OAI gNB. In OAI, the RACH preamble detector is implemented in the physical layer, where correlation-based detection and threshold comparison are performed.

\begin{table}[H]
    \centering
    \caption{RACH Parameters in Experiment}
        \begin{tabular}{ | c | c | c |} 
        \hline
        \rowcolor{lightgray} \textbf{Parameter} & \textbf{Component} & \textbf{Value} \\
        \hline 
        $ssb-PositionsInBurst$ & gNB & 1  \\ 
        \hline 
        $msg1-FDM$ & gNB & 1 \\ 
        \hline 
        $ssb-perRACH-Occasion$ & gNB & 1 \\ 
        \hline 
        $prach-configurationIndex$ & gNB & 160, 161 \\ 
        \hline
        $prach-configurationIndex$ & Attacker & 145, 147, 149 \\ 
        \hline
        $prach-configurationIndex$ & UE (from & 148, 149\\
        & gNB SIB1) & 148, 149\\
        \hline
        \end{tabular}
    \label{tab:RACHParameter}
\end{table} 

\begin{table}[H]
    \centering
    \caption{Experimental Parameters}
    \begin{tabularx}{\linewidth}{|l|X|l|} 
        \hline
        \rowcolor{lightgray} \textbf{Parameter} & \textbf{Description} & \textbf{Value} \\
        \hline 
        $p_{noise}$ & Background noise level on RACH & 17.4 dB \\ 
        \hline
        $p_{attacker}$ & Attacker Msg1 received power metric & 51 dB \\ 
        \hline
        $p_{UE}$ & UE Msg1 received power metric & 56.4 dB \\
        \hline
        $j$ & RO offset of attacker w.r.t. UE & 0..15 \\ 
        \hline
        $Band$ & Operating band & n78 \\
        \hline
        $f_{center}$ & Central frequency & 3619.20 MHz \\
        \hline
        $SCS$ & Subcarrier Spacing & 30 kHz \\
        \hline
        $BW$ & Bandwidth & 40 MHz \\
        \hline
        $G_{TX}$ & Antenna TX gain & 77.75 dB \\
        \hline
        $G_{RX}$ & Antenna RX gain & 58.00 dB \\
        \hline
    \end{tabularx}
    \label{tab:experimentParameter}
\end{table}

\subsection{Attack Strategies}

The attacker transmits Msg1 periodically with interval $T_a$ ROs. Three representative periods are tested:
\begin{itemize}
    \item $T_a = 1$: continuous flooding, representing a persistent jammer.  
    \item $T_a = 2$: moderate attacker, reducing duty cycle and energy cost.  
    \item $T_a = 16$: stealth/low-duty attacker, aiming for long-term disruption.  
\end{itemize}

These intervals cover a spectrum of attacker strategies from aggressive to stealthy. In addition, two gNB parameters are varied:
\begin{itemize}
    \item \textbf{Noise update factor $\beta$:} values of 0.24, 0.12, 0.06, and 0 are tested to analyze the adaptation speed of the threshold.  
    \item \textbf{Detection margin $\delta$:} varied to examine the sensitivity of gNB to discriminate legitimate Msg1 from noise.  
\end{itemize}

\subsection{Results and Validation}

\begin{figure}[h]
\centering
\includegraphics[width=0.39\textwidth]{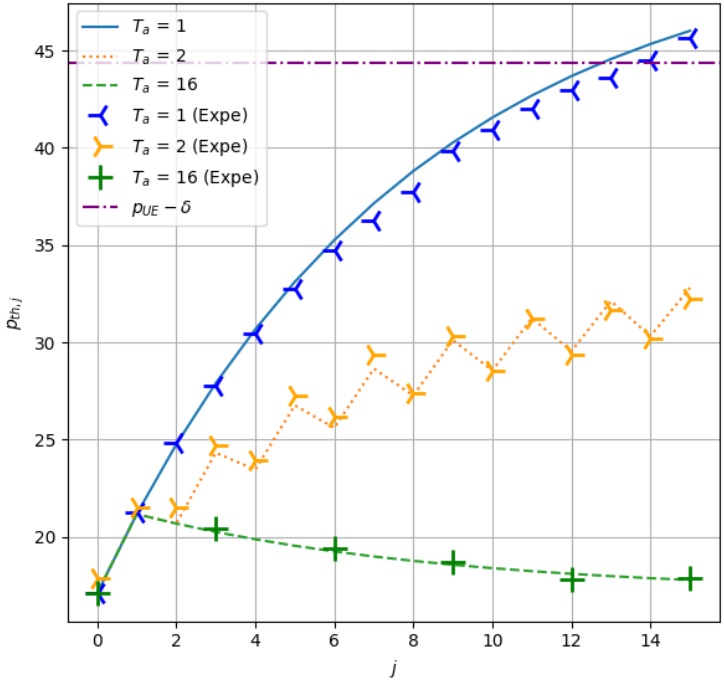}
\caption{$p_{th,j}$ for different $T_a$}
\label{fig:expe11noise}
\end{figure}

\begin{figure}[h]
\centering
\includegraphics[width=0.39\textwidth]{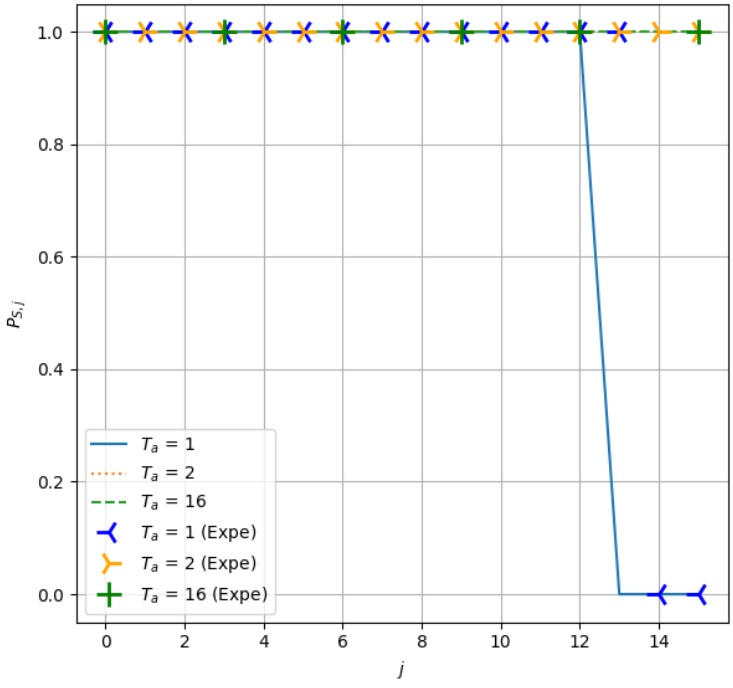}
\caption{$P_{S,j}$ for different $T_a$}
\label{fig:expe11success}
\end{figure}

Figures~\ref{fig:expe11noise} and ~\ref{fig:expe11success} shows the results of $p_{th,j}$ and $P_{S,j}$ for different values of $T_a$ in the experiment 1. The analytical model shown in the figures were derived in Section~\ref{sec:msg1Model}.  With $T_a=1$, the threshold rises rapidly and quickly surpasses the UE Msg1 power, blocking access. At $T_a=2$, accumulation is slower, while $T_a=16$ produces minimal threshold increase. These trends align with the model prediction that lower $T_a$ values lead to faster threshold buildup.  

\begin{figure}[h]
\centering
\includegraphics[width=0.39\textwidth]{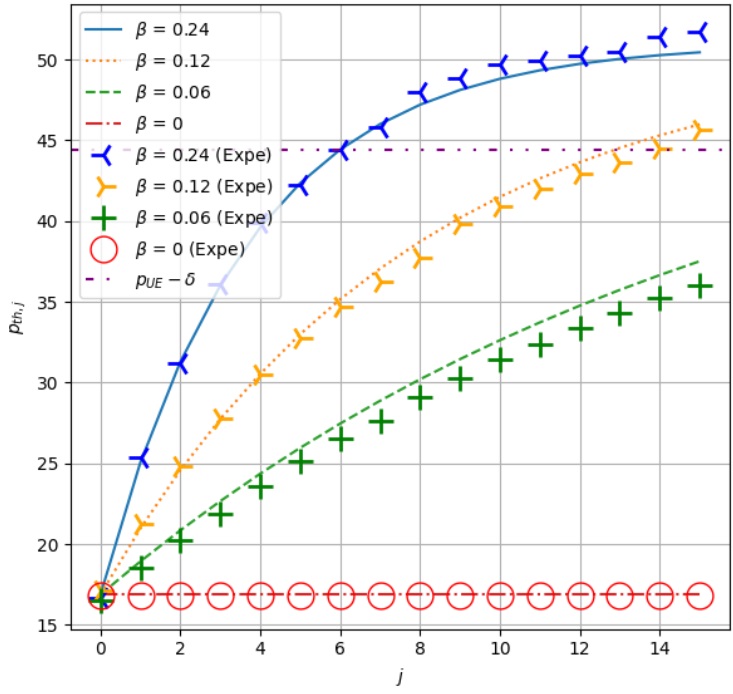}
    \caption{$p_{th,j}$ for different $\beta$}
    \label{fig:expe12noise}
\end{figure}

\begin{figure}[h]
\centering
\includegraphics[width=0.39\textwidth]{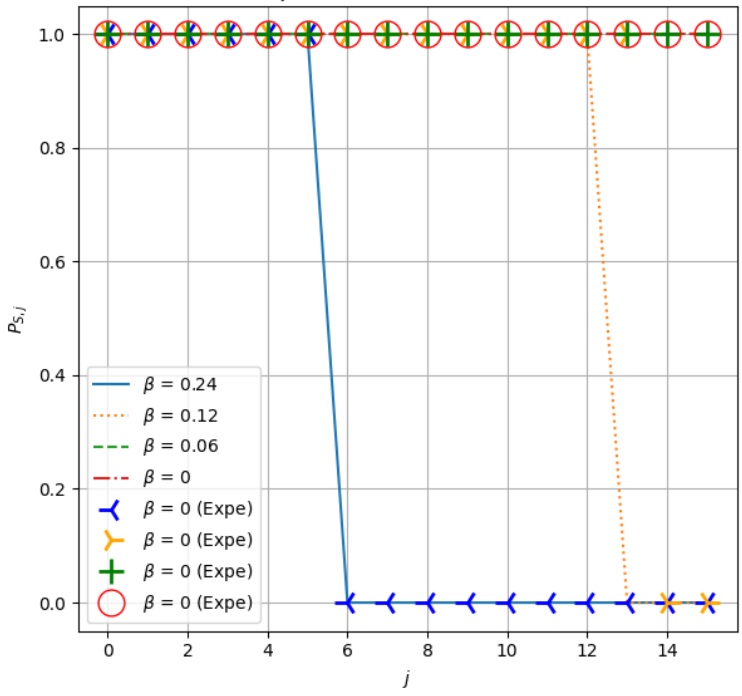}
    \caption{$P_{S,j}$ for different $\beta$}
    \label{fig:expe12success}
\end{figure}

Figures~\ref{fig:expe12noise} and ~\ref{fig:expe12success} show the results of $p_{th,j}$ and $P_{S,j}$ for different values of $\beta$ in the experiment 2, respectively. Smaller $\beta$ values slow down the threshold updates, allowing a higher probability of UE access success despite jamming. Setting $\beta=0$ makes the threshold static, which prevents attacker influence but also disables adaptation to real interference—similar to the behavior of srsRAN gNB defaults.  

\begin{figure}[h]
\centering
\includegraphics[width=0.39\textwidth]{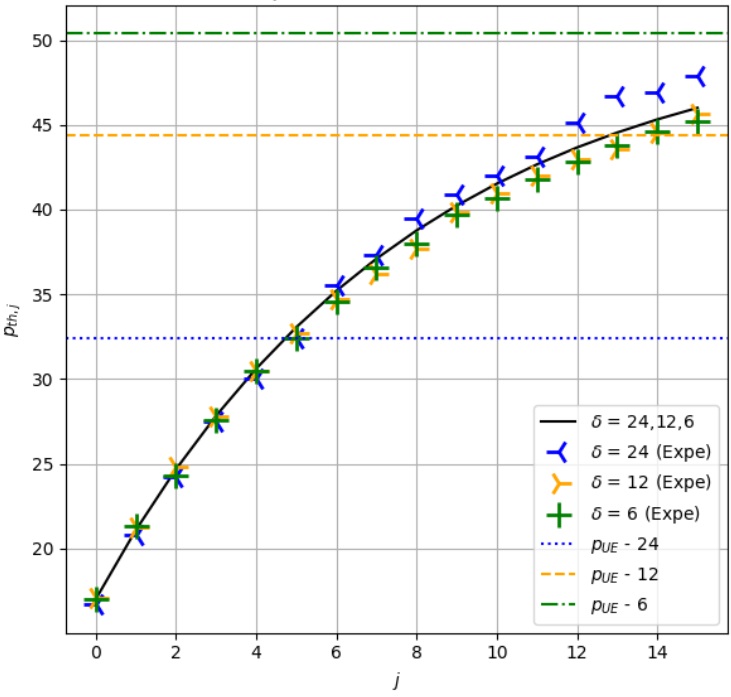}
    \caption{$p_{th,j}$ for different $\delta$ values}
    \label{fig:expe13noise}
\end{figure}

\begin{figure}[h]
\centering
\includegraphics[width=0.39\textwidth]{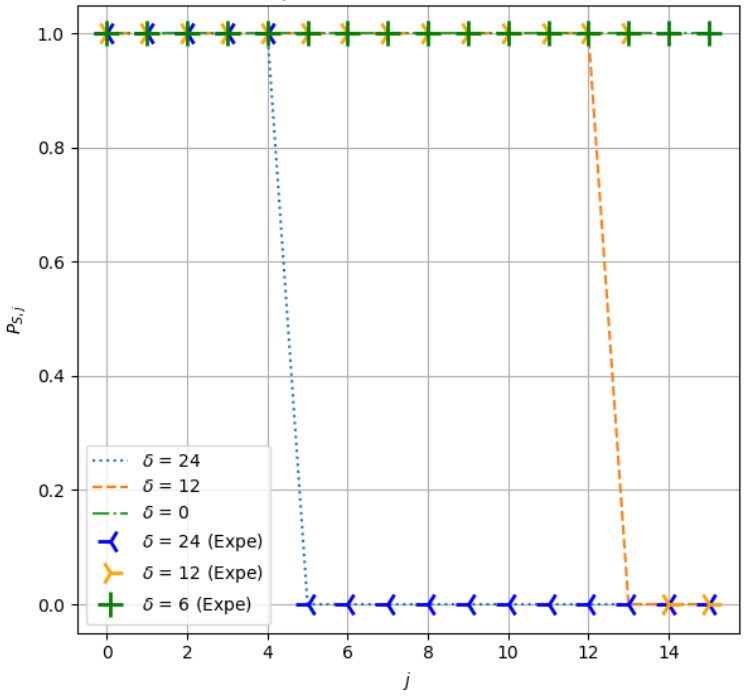}
    \caption{$P_{S,j}$ vs. $j$ for different $\delta$}
    \label{fig:expe13success}
\end{figure}

Figures~\ref{fig:expe13noise} and ~\ref{fig:expe13success} show $p_{th,j}$ and $P_{S,j}$ for different values of $\delta$ in the experiment 3. Smaller $\delta$ increases the access success probability of the legitimate UE by lowering the required differential above noise, but excessive sensitivity ($\delta=0$) may lead to false detections of noise as Msg1.  

\textbf{Experimental Insight:}  
\begin{itemize}
    \item Analytical predictions closely match experimental results for threshold evolution and access success probability.  
    \item Frequent attackers ($T_a=1$) are the most effective, while low-duty attackers ($T_a=16$) cause limited disruption.  
    \item Parameter tuning (lower $\beta$, optimized $\delta$) mitigates attack effects but introduces trade-offs in noise responsiveness and false alarm risk.  
\end{itemize}

We also verified the Msg1 attacker on an srsRAN gNB, which employs a memoryless threshold update ($\alpha=1, \beta=0, \gamma=0$) where only the most recent $p_{measured,i}$ is used. While such instantaneous updates simplify processing, they remain vulnerable to Msg1 jamming. When the attacker transmits a high-power Msg1 during an RO, the instantaneous estimate $p_{th,i}$ can spike above the UE’s Msg1 power, temporarily blocking access. Our srsRAN experiments confirmed this behavior: increasing $p_{attacker}$ sharply reduced $P_{S,j}$ as $p_{th,i}$ exceeded $p_{UE}-\delta$. This shows that even memoryless implementations are not immune to Msg1 jamming, but suffer from reduced access probability when exposed to periodic attacks. Regarding resilience, although one could argue that UEs might counteract by transmitting with higher power, in practice Msg1 is restricted by the maximum PRACH transmit power of the UE and is governed by open-loop PRACH power control. Therefore, under realistic conditions, the attacker remains effective, particularly when transmitting with short $T_a$.
These results validate the proposed analytical model and demonstrate how attacker periodicity and gNB parameter tuning jointly determine RACH resilience.

\section{Conclusion and Future Work}

This paper investigated Msg1-based RACH jamming in 5G networks through analytical modeling and experimental validation. The main findings are summarized as follows. Msg1 jamming raises the gNB's noise threshold, which potentially blocks legitimate UE access, particularly when the attacker transmits at short intervals. The proposed recursive model for the evolution of the noise threshold can accurately predict the impact of the attacker's periodicity and the gNB configuration parameters. The experimental results obtained from an OAI-based 5G testbed closely match the analytical predictions, confirming that even low-power Msg1 transmissions can significantly reduce the UE access probability.





In this study, the attacker can select only a single RACH occasion (RO) and generate only one preamble transmission at a time. One direction for future work is to implement an advanced attacker capable of simultaneously generating multiple preambles across multiple RACH occasions (ROs) in the frequency domain. Another direction is to investigate adaptive defense mechanisms to mitigate such attacks. These efforts will further enhance the understanding of RACH vulnerabilities and support the development of more resilient 5G/B5G access procedures.
\bibliographystyle{IEEEtran}
\bibliography{references}

\end{document}